\documentclass[aps,pra,superscriptaddress,reprint,showpacs,floatfix]{revtex4-1}

\usepackage{amsmath}
\usepackage{slashed}
\usepackage{txfonts}
\usepackage{microtype}
\usepackage{graphicx}
\usepackage[breaklinks=true]{hyperref}
\usepackage{color}

\def\sout{\bgroup\markoverwith
{\textcolor{red}{\rule[0.5ex]{2pt}{0.5pt}}}\ULon}
\def\be{\begin{equation}}
\def\ee{\end{equation}}
\def\bes{\begin{equation*}}
\def\ees{\end{equation*}}
\def\bea{\begin{eqnarray}}
\def\eea{\end{eqnarray}}
\def\beas{\begin{eqnarray*}}
\def\eeas{\end{eqnarray*}}
\def\bal#1\eal{\begin{align}#1\end{align}}
\def\bals#1\eals{\begin{align*}#1\end{align*}}
\newcommand{\bra}[1]{\langle #1|}
\newcommand{\ket}[1]{|#1\rangle}

\newcommand{\bk}[1]{\langle #1\rangle}

\renewcommand{\vec}{\vectorsym}

\renewcommand*{\vec}[1]{\boldsymbol{#1}}
\DeclareMathOperator*{\argmin}{arg\,min}
\DeclareMathOperator*{\argmax}{arg\,max}

\graphicspath{{figures/}}

\usepackage[normalem]{ulem}

\begin{document}


\title{Theory of the Rotating Polaron: Spectrum and Self-Localization}

\author{Enderalp Yakaboylu}
\affiliation{IST Austria (Institute of Science and Technology Austria), Am Campus 1, 3400 Klosterneuburg, Austria}
\author{Bikashkali Midya}
\affiliation{IST Austria (Institute of Science and Technology Austria), Am Campus 1, 3400 Klosterneuburg, Austria}
\affiliation{Department of Materials Science and Engineering, University of Pennsylvania, Philadelphia, PA, 19104, USA}
\author{Andreas Deuchert}
\affiliation{IST Austria (Institute of Science and Technology Austria), Am Campus 1, 3400 Klosterneuburg, Austria}
\author{Nikolai Leopold}
\affiliation{IST Austria (Institute of Science and Technology Austria), Am Campus 1, 3400 Klosterneuburg, Austria}
\author{Mikhail Lemeshko}
\affiliation{IST Austria (Institute of Science and Technology Austria), Am Campus 1, 3400 Klosterneuburg, Austria}

\date{\today}

\begin{abstract}

We study a quantum impurity possessing both translational and  internal rotational degrees of freedom interacting with a bosonic bath.  Such a system corresponds to a `rotating polaron', which  can be used to model, e.g., a rotating molecule immersed in an ultracold Bose gas or superfluid Helium. We derive the Hamiltonian of the rotating polaron and study its spectrum in the weak- and strong-coupling regimes using a combination of variational, diagrammatic, and mean-field approaches. We reveal how the coupling between linear and angular momenta affects stable quasiparticle states, and demonstrate that internal rotation leads to an enhanced self-localization in the translational degrees of freedom.

\end{abstract}

\maketitle

\section{Introduction}

Since the seminal papers of Landau and Pekar~\cite{landau1933uber,pekar1946local}, the polaron became one of the most studied models in condensed-matter physics. The polaron has been initially introduced as a quasiparticle consisting of an electron dressed by lattice excitations in a crystal, and can, in general, be considered as an elementary building block of complex condensed-matter systems. Over the years, polaron models have been adapted to several different context, and nowadays are used to describe more general classes of impurities interacting with a quantum many-particle bath~\cite{Chikkatur_2000,Schirotzek_09,Palzer_09,Spethmann_12,Scelle_13, Cetina_15,Jorgensen_16,Hu_16,Cetina96,fukuhara2013quantum,koschorreck2012attractive,kohstall2012metastability}.

In the conventional polaron problem, the impurity is considered as a structureless particle, such as an electron. However, there are several systems where additional internal degrees of freedom of the impurity cannot be neglected. A well-known example is molecules immersed in superfluid helium droplets, which have been used as a tool of molecular spectroscopy for over two decades, see e.g.\ Refs.~{\cite{toennies2004superfluid, StienkemeierJPB06, LemSchmidtChapter} and references therein. Although impurities with ``simple'' internal structure (e.g.\ spin$-1/2$)  have been actively studied since the works of Anderson and Kondo~\cite{Anderson_61,kondo1964resistance}, more complex degrees of freedom, such as rotational states of molecules, have not received as much attention from the community of condensed-matter physicists. Here, inspired by the recent advances in the polaron theory~\cite{frohlich1954electrons,appel1968solid,emin2013polarons,frohlich1963polarons,Pietro_14,Devreese15,grusdt2015renormalization,Grusdt_2015}  as well as by the recently introduced angulon quasiparticle (a quantum rotor dressed by a many-body field)~\cite{Lemeshko_2015, LemSchmidtChapter, PhysRevX.6.011012}, we establish a theory for the hybrid between the two species -- the `rotating polaron'  or `moving angulon'. More specifically, we consider rotating impurities, such as molecules, immersed in a   many-particle bosonic bath, and study stable and metastable states of the resulting quasiparticles. 

The paper is organized as follows. In Sec.~\ref{sec_hamiltonain},  we derive the Hamiltonian for a rotating polaron from first principles, and demonstrate the corresponding limits for the Fr\"{o}hlich polaron as well as the angulon Hamiltonians. In  Sec.~\ref{sec_weak}, we obtain the spectrum of the rotating polaron in the weak-coupling regime using a variational approach taking into account single-phonon excitations. In Sec.~\ref{sec_strong} we study the system in the strong-coupling regime using a Pekar-type ansatz, and discuss 
the self-localization transition in the rotating polaron. The conclusions of the paper are drawn  in Sec.~\ref{sec_conc}.

\section{The Hamiltonian} \label{sec_hamiltonain}

\begin{figure}[!b]
	\includegraphics[width=0.9\linewidth]{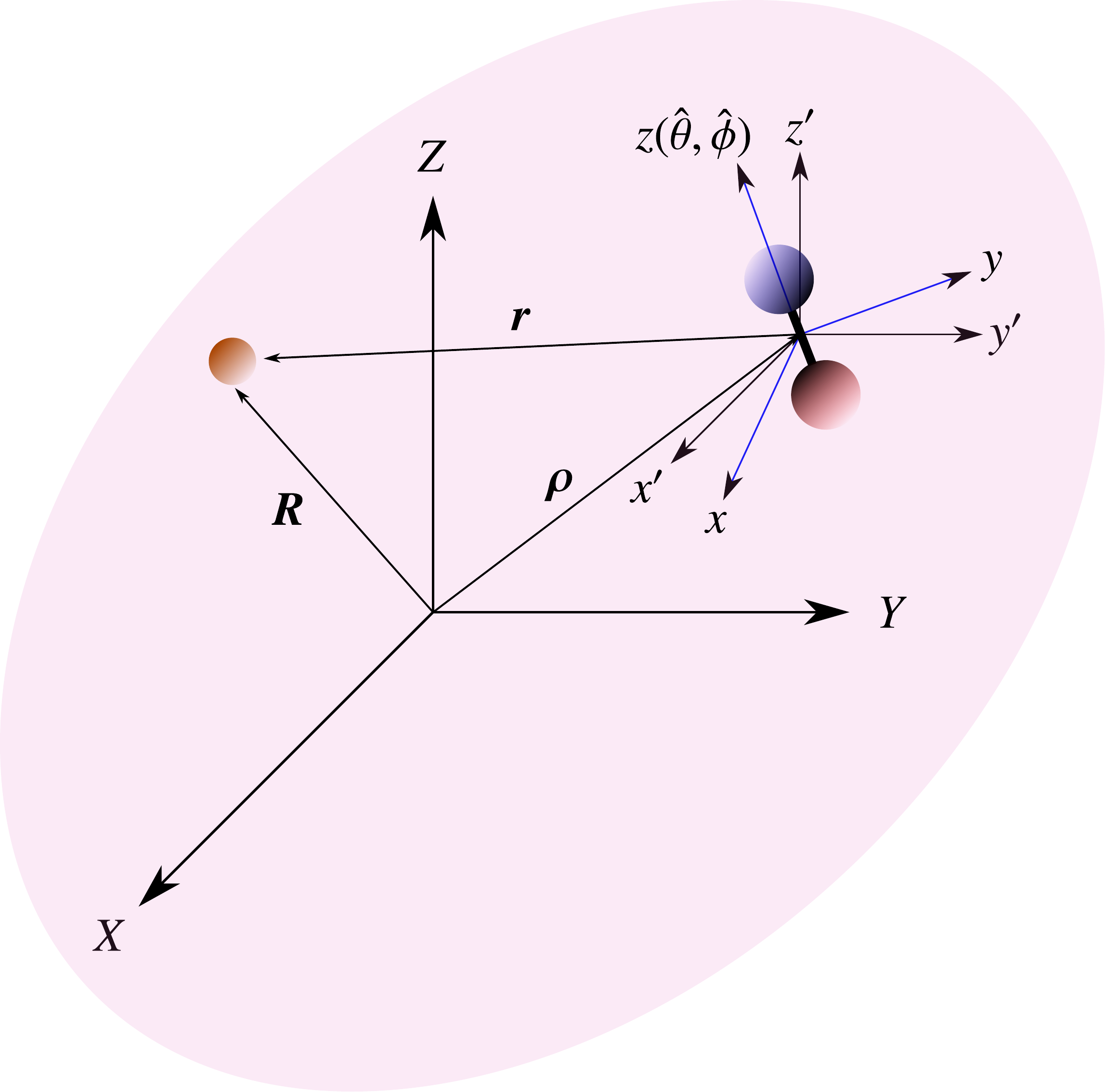} 
	\caption{Schematic diagram of a rotating impurity, whose center-of-mass (CM) is located at $\vec{\rho}$, interacting with a boson at $\vec{R}$. Here $(XYZ), (x',y',z')$, and $(x,y,z)$ are the laboratory, CM, and body-fixed coordinate frames, respectively. See the text.}
\label{diagram}
\end{figure}

The Hamiltonian that describes a mobile impurity with rotational degrees of freedom immersed in a bosonic bath is given by:
\be
\label{ham}
\hat H = \frac{\vec{\hat{P}}^2}{2M} + B \vec{\hat{L}}^2 + \sum_{\vec{k}} \omega(k) \hat b^\dagger_{\vec{k}} \hat b_{\vec{k}} +\hat{H}_\text{int} \,
\ee
(in what follows we use the units of $\hbar \equiv 1$). Here the first two terms represent the total translational and rotational kinetic energy of the extended impurity with mass $M$ and rotational constant $B = 1/(2I)$, where $I$ is the impurity's moment of inertia. The third term, with  $\sum_{\vec{k}} \equiv \int d^3 k / (2 \pi)^3$, corresponds to the kinetic energy of the bosons parametrised by the dispersion relation, $\omega(k)$. The bosonic creation and annihilation operators, $\hat b^\dagger_{\vec{k}}$ and $\hat b_{\vec{k}}$, obey the commutation relation $[\hat b_{\vec{k}}, \hat b^\dagger_{\vec{k}'}] = (2 \pi)^3 \delta (\vec{k} - \vec{k}')$. Finally, the last term is the impurity-bath interaction. 

In the Bogoliubov approximation~\cite{LemSchmidtChapter}, the interaction Hamiltonian can be written as  
\be
\label{H_int}
\hat{H}_{\rm int} = \sqrt{n} \sum\limits_{\bf k} \sqrt{\frac{\epsilon(k)}{\omega(k)}} e^{-i{\bf k}\cdot\hat{\boldsymbol\rho}} V({\bf k},\hat{\theta},\hat{\phi}) \hat{b}^\dag_{\bf k} + \text{H.c.}\, .
\ee
A very similar Hamiltonian can be obtained for a crystalline solid, by performing an expansion in atomic displacements and keeping only linear terms. In Eq.~\eqref{H_int}, $\epsilon(k) = k^2/2m_b$ is the kinetic energy of each atom  with the mass $m_b$ in the bath, and $n$ is the particle density. Furthermore, $\exp(-i \vec{k} \cdot \hat{\vec{\rho}})$  is the Fourier-transformed density of an impurity, where the operator $\hat{\vec{\rho}}\equiv (\hat{\rho},{\hat\theta}_\rho,\hat{\phi}_\rho)$, which is conjugate to $\vec{\hat{P}}$, measures an instantaneous position of the molecule's center-of-mass (CM) frame with respect to the laboratory frame, as schematically illustrated in Fig~\ref{diagram}. $ V({\bf k},\hat{\theta},\hat{\phi}) $ is the impurity-boson potential in Fourier space, where  the angle operators, $(\hat{\theta}, \hat{\phi})$,  describe the molecular orientation in the CM frame. After we expand the bosonic creation and annihilation operators, the plane wave $\exp(-i \vec{k} \cdot \hat{\vec{\rho}})$, and $V({\bf k},\hat{\theta},\hat{\phi}) $ in the spherical basis, we obtain (see Appendix~\ref{appendix}):
\be
\label{int1}
 \hat H_\text{int}  =  \sum_{ k \lambda \mu} \sum_{\ell \delta \alpha \gamma} \mathcal{U}_{\ell\alpha\lambda}^{\delta\gamma\mu}(k)~j_\ell(k \hat{\rho})  ~Y_{\ell \delta}^*(\hat{\Omega}_\rho)~Y_{\alpha \gamma}^*(\hat{\Omega}) ~\hat{b}^\dag_{k\lambda \mu}  + \text{H.c.} 
\ee
Here $\sum_k \equiv \int_0^\infty d k$, $\Omega \equiv (\theta,\phi)$, $j_\ell(k \hat{\rho}) $ is the spherical Bessel function of the first kind, and $~Y_{\ell \delta}(\hat{\Omega}_\rho)$ are the spherical harmonics. The coupling term, $ \mathcal{U}_{\ell\alpha\lambda}^{\delta\gamma\mu}(k) = U_\alpha (k) \sqrt{4\pi(2\alpha+1)(2l+1)/(2\lambda+1)} i^{\lambda-\alpha-l} C^{\lambda 0}_{\alpha 0, l 0} C^{\lambda \mu}_{\alpha \gamma, l \delta} \, $, with $C^{\lambda \mu}_{\alpha \gamma, l \delta}$ being the Clebsch-Gordan coefficients~\cite{Varshalovich}, couples both the impurity's CM translational motion and its internal rotation to many-particle excitations. $U_\alpha (k)$ is the angular-momentum-dependent coupling strength which we define in Eq.~(\ref{u_lambda}) below.

The Hamiltonian~(\ref{ham}) features both translational and rotational symmetry. Translational symmetry follows from the fact that the total linear momentum of the system, 
\be
\label{total_p}
\hat{\vec{\Pi}} = \hat{\vec{P}} + \sum_{\vec{k}} \vec{k} \hat b^\dagger_{\vec{k}} \hat b_{\vec{k}} \, ,
\ee
commutes with the Hamiltonian~(\ref{ham}) (this can be seen from Eq.~(\ref{H_int}) below). In Eq.~(\ref{total_p}), the second term is the collective linear momentum of the many-particle bath. The total angular momentum of the system, on the other hand, can be written as
\be
\label{total_ang_mom}
\vec{\hat{J}} = \hat{\vec{L}} + \hat{\vec{L}}_T + \sum_{k \lambda \mu \nu} \vec{\sigma}^\lambda_{\mu \nu} \hat b^\dagger_{k \lambda \mu}  \hat b_{k \lambda \nu} \, ,
\ee
where $\hat{\vec{L}}_T$ is the corresponding angular momentum of the translational motion stemming from the spherical decomposition of $\hat{\vec{P}}^2 = \hat{P_\rho}^2 + \hat{\vec{L}}_T^2/\hat{\rho}^2$, with $\hat{P_\rho}^2$ the radial part.  The last term of Eq.~(\ref{total_ang_mom}) is the collective angular momentum of the bath with $\vec{\sigma}^\lambda_{\mu \nu}$ being the spin-$\lambda$ representation of the SO(3) group. The Clebsch-Gordan coefficient, $C^{\lambda \mu}_{\alpha \gamma, l \delta}$, in the coupling $\mathcal{U}_{\ell\alpha\lambda}^{\delta\gamma\mu}(k)$ ensures that the total angular momentum operator commutes with the Hamiltonian~(\ref{ham}), which is therefore rotationally invariant. 

We note, however, that the total angular momentum operator does not commute with the total linear momentum operator, i.e., $[\vec{\hat{J}}^2,\vec{\hat{\Pi}}] \neq 0 \neq [\hat{J}_z,\vec{\hat{\Pi}}]$. Therefore, the state of the combined impurity-bath system can be specified either by the eigenvalues of $\hat H$ and $\hat{\vec{\Pi}}$ or by the eigenvalues of $\hat{H}$, $\vec{\hat{J}}^2$, and $\hat{J}_z$. For convenience, the state defined by the former set of eigenvalues can be called `the rotating polaron', while the latter one can be referred to as `the moving angulon' -- the quasiparticle defined through the total angular momentum.

By using the translation operator, 
\be
\hat T(\hat{\vec{\rho}})  = \exp\left[ - i \hat{\vec{\rho}} \cdot \sum_{\vec{k}} \vec{k} \hat b^\dagger_{\vec{k}} \hat b_{\vec{k}} \right] \,, 
\ee
the interaction Hamiltonian~(\ref{int1}) can be written in a more elegant way as
\be
\label{H_int}
\hat H_\text{int} = \hat T(\hat{\rho}) \left( \sum_{k \lambda \mu} U_\lambda (k) Y_{\lambda \mu}^* (\hat \Omega) \hat b^\dagger_{k \lambda \mu}  + \text{H.c.}  \right) \hat T^{-1} (\hat{\rho}) \, ,
\ee 
where
\be
\label{u_lambda}
U_\lambda (k)=  \sqrt{\frac{8 n k^2 \epsilon (k)}{\omega (k) (2\lambda+1)}} \int_0^\infty dr r^2 V_\lambda (r) j_\lambda (k r) \, 
\ee
is the angular-momentum-dependent coupling strength. In Eq.~(\ref{u_lambda}), $V_\lambda (r) $ is the interaction potential in the angular momentum channel $\lambda$ defined via the expansion of the corresponding impurity-boson potential in the body-fixed coordinates, $V({\bf r}) = \sum_\alpha V_\alpha(r) Y_{\alpha 0}(\theta_r,\phi_r) $, which is chosen to be rotationally symmetric around the $z$-axis. In the translated frame, the interaction Hamiltonian between the impurity and the bath depends solely on the molecular orientation and not on the CM coordinate $\vec{\rho}$. Therefore, the interaction is a  result of the angular momentum transfer in the system. As we discuss below, this corresponds to the angulon Hamiltonian.

\subsection{The angulon limit}

The form of the interaction Hamiltonian~(\ref{H_int}) suggests that the Hamiltonian in the translated frame simply reads
\be
\label{ham1}
\hat H'  \equiv \hat T^{-1} \hat H  \hat T  = \hat H_\text{ang} + \frac{1}{2M} \left(\vec{\hat{P}} -  \sum_{\vec{k}} \vec{k} \hat b^\dagger_{\vec{k}} \hat b_{\vec{k}}\right)^2  \,.
\ee
Here the first term is the so-called angulon Hamiltonian, which describes a rotating impurity interacting with a bath of bosons~\cite{Lemeshko_2015}:
\be
\hat H_\text{ang} = B \vec{\hat{L}}^2 + \sum_{k \lambda \mu} \omega(k )\hat b^\dagger_{k \lambda \mu} \hat b_{k \lambda \mu} + \sum_{k \lambda \mu} U_\lambda (k) Y_{\lambda \mu}^* (\hat\Omega) \hat b^\dagger_{k \lambda \mu}  + \text{H.c.} \, 
\ee
Recently, it has been demonstrated that impurities whose orbital angular momentum is coupled to a many-particle bath form the so-called `angulon quasiparticles'~\cite{Lemeshko_2015, LemSchmidtChapter, PhysRevX.6.011012}. This  quasiparticle can be thought of as a non-Abelian counterpart of the polaron~\cite{Devreese15}, as it represents a quantum rotor dressed by a many-body bosonic field. Moreover it was demonstrated that the predictions of the angulon theory are in good agreement with experiments on molecules embedded in superfluid helium nanodroplets~\cite{lemeshko2016quasiparticle,Shepperson16,Cherepanov}.

The physical meaning of the above decomposition is quite transparent. When $M \to \infty$, one can neglect the center-of-mass motion of a molecule such that the Hamiltonian~(\ref{ham1}) reduces to the angulon Hamiltonian. Furthermore, the decomposition~(\ref{ham1}) is very useful for a perturbative analysis of a moving angulon with large mass, which can be described as a rotating impurity whose translational motion is perturbed by the many-particle bath only weakly.

\subsection{The polaron limit}

The spherically symmetric part of the angular-momentum-dependent coupling, $U_0 (k)$, in fact, defines the coupling between the impurity's CM motion and the many-particle bath. Accordingly, if we define $V_F (k)$ through $ U_0 (k) =  V_F (k) \sqrt{2 k^2/\pi}$, we can further decompose the transformed Hamiltonian~(\ref{ham1}) into
\be
\label{ham2}
\hat H' = \hat H_\text{pol} + B \vec{\hat{L}}^2 + \sum_{k \lambda \neq 0 \mu} U_\lambda (k) Y_{\lambda \mu}^* (\hat \Omega) \hat b^\dagger_{k \lambda \mu}  + \text{H.c.} \, , 
\ee
where 
\bal
\label{polaron_ham}
\nonumber \hat H_\text{pol} & = \frac{1}{2M} \left(\vec{\hat{P}} -  \sum_{\vec{k}} \vec{k} \hat b^\dagger_{\vec{k}} \hat b_{\vec{k}}\right)^2 + \sum_{\vec{k}} \omega(k) \hat b^\dagger_{\vec{k}} \hat b_{\vec{k}} \\
& + \sum_{\vec{k}} V_F(k) \left( \hat b^\dagger_{\vec{k}} + \hat b_{\vec{k}} \right) \, ,
\eal
is the Fr\"{o}hlich polaron Hamiltonian written in the Lee-Low-Pines (LLP) form~\cite{LLP_53}.

For a structureless spherically symmetric impurity, the taking the limit of $B \to 0$ and $U_{\lambda \neq 0 }(k) \to 0$ reduces the Hamiltonian~(\ref{ham2}) to the polaron Hamiltonian~(\ref{polaron_ham}). Moreover, for moving impurities whose rotational coupling to the bath is small compared to the translational coupling, one can account for the angular part of the Hamiltonian~(\ref{ham2}) perturbatively.

\section{The Weak-Coupling Regime} \label{sec_weak}

\subsection{The Variational Approach}

In what follows, we study the problem non-perturbatively by means of a variational ansatz. Our goal is to investigate the rotating polaron, that is, we consider a quasiparticle labeled by the total linear momentum of the impurity-bath system. This can be done most conveniently in the translated frame, where the total linear momentum operator is simply given by $\hat{\vec{P}} = \hat T^{-1} \hat{\vec{\Pi}}  \hat T $. In this case, we can replace the operator $\hat{\vec{P}}$ in the Hamiltonian~(\ref{ham1}) with the corresponding eigenvalue, $\vec{p}$, which defines the total linear momentum of the impurity-bath system in the laboratory frame. Accordingly, we introduce the following trial wavefunction for the rotating polaron:
\be
\ket{\Psi_{\vec{p}}} = \left( g \ket{j m} \ket{0} + \sum_{\vec{k}' j' m'} \alpha_{j'm'} (\vec{k}') \ket{j'm'} \hat b^\dagger_{\vec{k}'} \ket{0}\right) \ket{\vec{p}} \, ,
\ee
which corresponds to the expansion of the quasiparticle state over zero- and one-phonon bath excitations. Therefore, it is supposed to be a good approximation in the weak-coupling regime. The corresponding trial state in the laboratory frame reads
\be
\label{t_var}
\hat T \ket{\Psi_{\vec{p}}} = g \ket{j m} \ket{0} \ket{\vec{p}} + \sum_{\vec{k}' j' m'} \alpha_{j'm'} (\vec{k}') \ket{j'm'} \hat b^\dagger_{\vec{k}'} \ket{0} \ket{\vec{p} - \vec{k}'} \, .
\ee
Here, the first part of the trial state describes the impurity which is moving with linear momentum $\vec{p}$, and rotating with the angular quantum numbers $j$ and $m$ in the absence of phonons. The second term of the trial state~\eqref{t_var} corresponds to the impurity + one-phonon state with the total linear momentum $\vec{p}$. Since we consider the rotating polaron, and the linear momentum operator does not commute with the angular momentum operator, the total angular momentum of this second term cannot be specified. On the other hand, this angular momentum can be specified in the moving angulon picture (where the linear momentum is, in turn, not defined), see the discussion above. We further note that it is more convenient to describe the moving angulon in the labarotary frame, as the total angular momentum operator in the translated frame, $\vec{\hat{J}}' = \hat T^{-1} \vec{\hat{J}}  \hat T  $, is quite involved.

The minimization of the functional $\bra{\Psi_{\vec{p}}} \hat{H}' - E \ket{\Psi_{\vec{p}}}$ with respect to the variational parameters, $g^*$ and $\alpha_{j m}^{*} (\vec{k})$, leads to the following expression for the variational energy:
\be
\label{energy_ap}
E = \frac{\vec{p}^2}{2M} + B j (j+1) - \Sigma_{jm \vec{p}}(E) \, ,
\ee
where the self-energy is given by
\be
\Sigma_{jm \vec{p}}(E) = \sum_{j'm' \vec{k}'} \frac{\left|\bra{jm} \bra{0} \left( \sum_{k \lambda \mu} U_\lambda (k) Y_{\lambda \mu}(\hat \Omega) \hat b_{k \lambda \mu}\right)\ket{j'm'} \hat b^\dagger_{\vec{k}'} \ket{0} \right|^2}{(\vec{p}-\vec{k}')^2/(2M) + B j' (j'+1) + \omega(k) -  E} \, .
\ee
Here the free quantum numbers, $j$ and $m$, label the corresponding angular momentum numbers of the impurity, whereas the quantum number $\vec{p}$ stands for the total linear momentum of the impurity-bath system, as discussed above.

In the iterative solution to Eq.~\eqref{energy_ap}, the leading-order term is given by $E^{(1)} = \vec{p}^2/(2M) + B j (j+1)$, and the second-order term reads $E^{(2)}= \vec{p}^2/(2M) + B j (j+1) - \Sigma_{jm \vec{p}}(E^{(1)})$, which matches the second order perturbation theory. Therefore, the variational energy~(\ref{energy_ap}) goes beyond the perturbative result, and hence corresponds to a resummation over all diagrams describing single-phonon excitations, see Refs.~\cite{LemSchmidtChapter, Bighin_2017} for further details.

The self-energy can be expressed as
\begin{widetext}
\be
\label{self_energy_ap}
\Sigma_{jm \vec{p}}(E) = \sum_{k l j' \lambda \lambda'} \frac{U_\lambda(k)U_{\lambda'}(k) (2\lambda +1)}{4\pi} \sqrt{\frac{(2l+1)(2\lambda'+1)(2j+1)}{4\pi}}(-1)^{j'+j+\lambda'}i^{\lambda-\lambda'} C_{\lambda 0, j 0}^{j'0}C_{\lambda' 0, j 0}^{j'0}C_{l 0, \lambda 0}^{\lambda' 0}C_{jm, l 0}^{jm} \begin{Bmatrix}
    \lambda & j' & j\\
    j & l & \lambda'
 \end{Bmatrix} d_{l0} \, ,
\ee
\end{widetext}
where $\{ \cdots \}$ is the  $6j$ symbol~\cite{Varshalovich}, and
\be
d_{l0} = \int  \frac{d \Omega_k Y_{l0}(\Omega_k)}{(\vec{p}-\vec{k})^2/(2M) + B j' (j'+1) + \omega(k) -  E} \, .
\ee
We would like to emphasize that in the limit of $M\to \infty$, the coefficient $d_{l0}$ is given by 
\be
d_{l0} = \frac{\sqrt{4\pi} \delta_{l0}}{ B j' (j'+1) + \omega(k) -  E} \, ,
\ee
such that the self-energy~(\ref{self_energy_ap}) reduces to the angulon self-energy reported in Ref.~\cite{Lemeshko_2015}. In such a case, the quantum numbers $j$ and $m$ define the angular momentum numbers of the angulon quasiparticle.

Similarly, in the other limit given by $B \to 0$ and $U_{\lambda \neq 0 }(k) \to 0$, the extended impurity shrinks to a structureless spherically symmetric particle, and we obtain
\be
\Sigma_{\vec{p}}(E) = \sum_{\vec{k}} \frac{V_F(k)^2}{(\vec{p}-\vec{k})^2/(2M) + \omega(k) -  E} \, ,
\ee
which coincides with the self-energy of the Fr\"{o}hlich polaron in the weak-coupling regime~\cite{Devreese15}, also see Ref.~\cite{Li_2014} for the many-impurity case.

In general, the energy~(\ref{energy_ap}) is found self-consistently via the poles of the Green's function, 
\be
G_{jm \vec{p}} (E) = \frac{1}{\vec{p}^2/(2M) + B j (j+1) - \Sigma_{jm \vec{p}} (E) - E} \, ,
\ee
which leads to Eq.~\eqref{energy_ap}. The entire excitation spectrum of the system is captured by the spectral function $\mathcal{A}_{jm \vec{p}} = \text{Im}\left[G_{jm \vec{p}} (E + i 0^+)\right]$.

\subsection{The Spectrum}

For the sake of consistency with other angulon studies, here we adapt the parameters used in Ref.~\cite{Lemeshko_2015}, which reproduce the order of magnitude of the relevant quantities for a molecule immersed in a Bose-Einstein condensate. Namely, we consider a bath with the Bogoliubov dispersion relation, $\omega (k) = \sqrt{\epsilon (k) (\epsilon (k) + 2 g_{\text{bb}} n)}$~\cite{Pitaevskii2016}, where $g_{\text{bb}} = 4\pi a_{\text{bb}}/m_b$, and we set the boson-boson scattering length to $a_{\text{bb}} = 3.3 /\sqrt{m_b B}$. We choose the impurity-boson interaction as $U_\lambda (k) = \sqrt{8 n k^2 \epsilon (k) /(\omega (k) (2\lambda+1))} \int dr r^2 V_\lambda (r) j_\lambda (k r)$. The coupling is modeled using Gaussian functions, $V_\lambda (r) =  u_\lambda (2\pi)^{-3/2} e^{-r^2/(2 r_\lambda^2)}$, taking into account the  leading orders, $\lambda =0,1$, and setting the parameters to $u_0 = 1.75 u_1 = 218 B$, and $r_0 = r_1 = 1.5 /\sqrt{m_b B}$. 

\begin{figure}
  \centering
  \includegraphics[width=\linewidth]{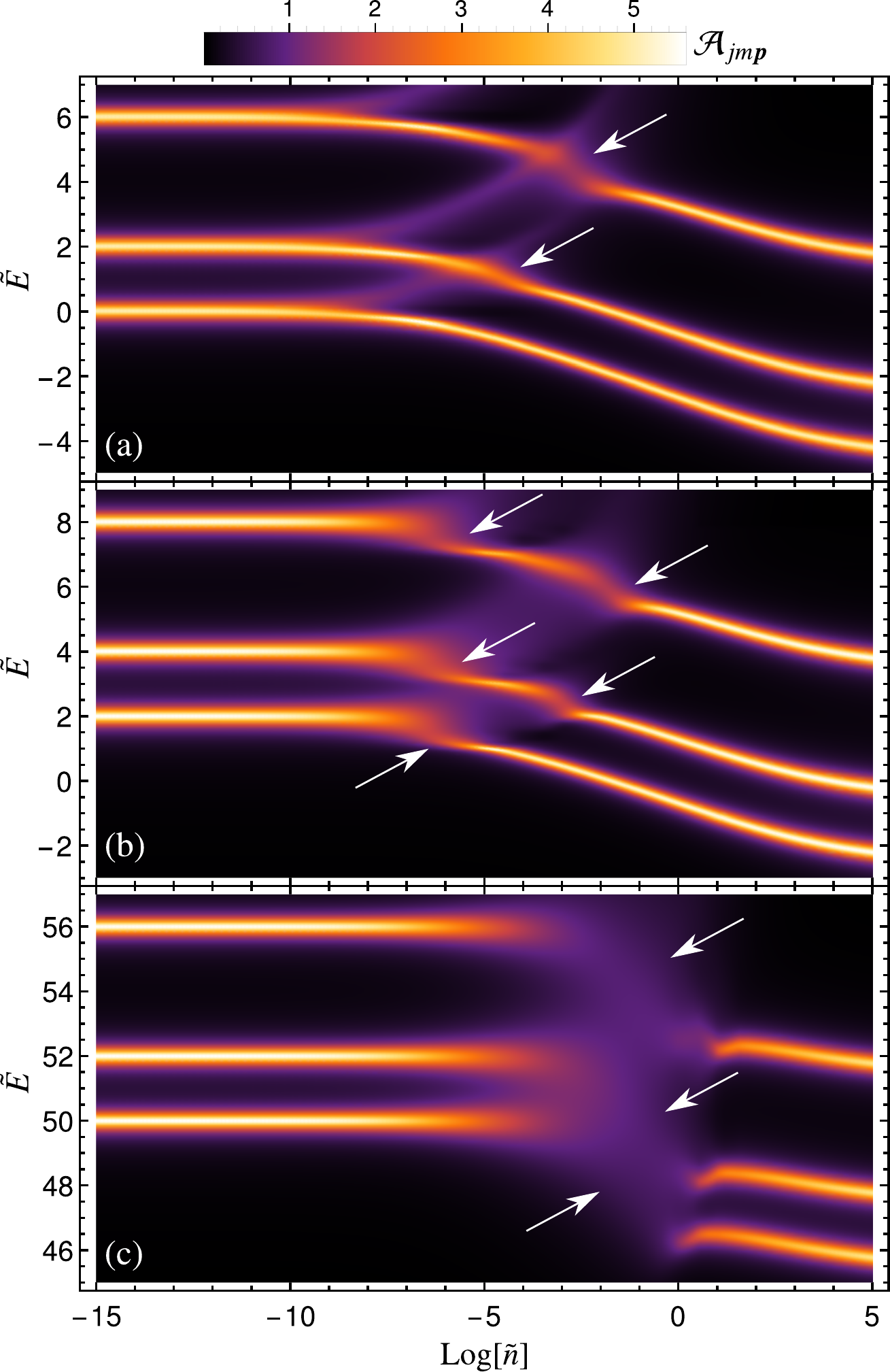}
 \caption{The spectral function of the rotating polaron, $\mathcal{A}_{j m \vec{p}}$, for the lowest three angular momentum states, $j=0,1,2$, with $m=0$, as a function of the dimensionless bath density, $\tilde{n} = n (m_b B)^{3/2}$, and the energy $\tilde{E} = E/B$. The cases shown correspond to the following values of total  linear momentum: (a)~$\vec{p} = 0$ (the angulon case), (b)~$\vec{p} = 2 \sqrt{M B} \hat{\vec{z}}$, and (c) $\vec{p} = 10 \sqrt{M B} \hat{\vec{z}}$. Instabilities are indicated by white arrows. See the text.}
 \label{var}
\end{figure}

In Fig.~\ref{var}, we plot the spectral function $\mathcal{A}_{jm \vec{p}}$ of a rotating polaron for the lowest three states, $j=0,1,2$, with $m=0$,  as a function of energy and bath density, for different values of linear momentum. Sharp light features in the figure correspond to long-lived quasiparticle states, while blurred red and purple regions correspond to metastable states with shorter lifetimes. In Fig.~\ref{var}~(a) we present the angulon case, which is given by $\vec{p} = 0$, and the spectrum naturally coincides with that obtained in Ref.~\cite{Lemeshko_2015}. As discussed in Ref.~\cite{Lemeshko_2015}, at the vicinity of metastable states, an angular momentum exchange between the impurity and the bath takes place, which corresponds to the so-called `angulon instabilities'~\cite{Cherepanov}} indicated in Fig.~\ref{var} by arrows. In Fig.~\ref{var}~(b), we consider the system with the total linear momentum $\vec{p} = 2 \sqrt{M B} \hat{\vec{z}}$. We find that for the excited states with $j=1,2$, there are two instability regimes. Furthermore, we uncover an instability for the ground state, $j=0$, which is absent in the angulon case. From the latter observation, we deduce that the instability arising in the ground state is a consequence of angular momentum exchange between the bath and the translational degree of freedom of the impurity. This further explains two instabilities observed in the excited states: the first one is a consequence of the exchange between the bath and the internal rotational degree of freedom (like in the angulon case), and the second one is a result of coupling between the bath and the impurity's translational degree of freedom. Moreover, if we further increase the linear momentum, the number of instabilities grows as a result of a resonant angular momentum transfer, and at some point they merge into a single broader instability. In Fig.~(\ref{var})~(c) we show the case of $\vec{p} = 10 \sqrt{M B} \hat{\vec{z}}$, where such a broad instability  leads to a break down of the quasiparticle picture for the bath densities of $-5~\lesssim~\text{Log}[\tilde{n}]~\lesssim~1$.

\section{The strong coupling regime and  Self-localization} \label{sec_strong}

\subsection{The Pekar Ansatz}

As the next step, we study the regime where the coupling between the impurity and the bath is strong. Such a regime can be approached using a mean-field Pekar-type ansatz~\cite{pekar1946local,Devreese15}. For this purpose, we first rewrite the Hamiltonian~(\ref{ham}) in the following form:
\bal
\label{ham3}
\hat H & = \frac{\vec{\hat{P}}^2}{2M} + B \vec{\hat{L}}^2 + \sum_{\vec{k}} \omega(k) \hat b^\dagger_{\vec{k}} \hat b_{\vec{k}} \\
\nonumber & + \sum_{\vec{k}} \sum_{\lambda \mu} (2\pi)^{3/2} i^{-\lambda} \frac{U_\lambda (k) Y_{\lambda \mu}(\Omega_k)}{k} Y_{\lambda \mu}^* (\hat \Omega) e^{- i \vec{k} \cdot \hat{\vec{\rho}}} \hat b^\dagger_{\vec{k}} + \text{H.c.} \, ,
\eal
where we have used the Cartesian representation of the boson operators, $\hat b^\dagger_{\vec{k}}$, see Appendix~\ref{appendix}.

Next, we introduce the Pekar ansatz,
\be
\ket{\Psi_P} = \ket{\varphi_I} \otimes \ket{\xi_B} \,,
\ee
where $\ket{\varphi_I}$ and $\ket{\xi_B}$  correspond to the impurity wavefunction and the bosonic state, respectively. After taking the expectation value, $\bra{\varphi_I} \hat H \ket{\varphi_I}$, the resulting effective bosonic Hamiltonian can be diagonalized using the following coherent-state transformation:
\be
\hat U = \exp\left[-\sum_{\vec{k}} \frac{1}{\omega(k)}\left( \tilde{V}(\vec{k}) \hat b^\dagger_{\vec{k}}- \text{H.c.}\right) \right] \, ,
\ee
where 
\be
\tilde{V}(\vec{k}) = \sum_{\lambda \mu} (2\pi)^{3/2} \frac{i^{-\lambda} U_\lambda(k) Y_{\lambda \mu} (\Omega_k)}{k} \bk{Y_{\lambda \mu}^{*} (\hat \Omega) e^{-i \vec{k}\cdot \hat{\vec{\rho}}}}_I \, ,
\ee
and $\bk{\hat A}_I \equiv \bra{\varphi_I} \hat A \ket{\varphi_I}$. The bosonic state, which minimizes the Pekar energy, is given by $ \ket{\xi_B} = \hat U \ket{0} $. Thereby, the respective energy yields:
\be
\label{pekar_energy}
\varepsilon_0 = \frac{1}{2M} \bk{\vec{\hat{P}}^2}_I + B \bk{\vec{\hat{L}}^2}_I - \sum_{\vec{k}} \frac{|\tilde{V}(\vec{k})|^2}{\omega(k)} \, ,
\ee
where the last term corresponds to the deformation energy of the bath in the limit of $M\to \infty$ and $B \to 0$. The ground state energy~(\ref{pekar_energy}) can also be written in terms of the Pekar energy functional as
\bal
\label{pekar_functional}
& \varepsilon_0 [\varphi_I]  = \int d^3 \rho \, d \Omega \, \left( \frac{1}{2 M} |\vec{\nabla}_{\vec{\rho}} \varphi_I (\vec{\rho},\Omega)|^2 +  B |\vec{\nabla}_\Omega\varphi_I (\vec{\rho},\Omega)|^2 \right) \\
\nonumber &- \int d^3 \rho \, d^3 \rho' \,  d \Omega\, d \Omega' |\varphi_I(\vec{\rho},\Omega)|^2 |\varphi_I(\vec{\rho'},\Omega')|^2 \, \tilde{U}(\vec{\rho},\vec{\rho}',\Omega,\Omega') \, ,
\eal
where
\bal
\label{self_potential}
\tilde{U}(\vec{\rho},\vec{\rho}',\Omega,\Omega') & = \sum_{\vec{k}} \sum_{\lambda \lambda' \mu \mu'} \frac{(2\pi)^3}{\omega(k) k^2} i^{\lambda'-\lambda}U_\lambda(k)U_{\lambda'}^{*}(k) \\
\nonumber & \times Y_{\lambda \mu} (\Omega_k)Y_{\lambda' \mu'}^* (\Omega_k) Y_{\lambda \mu} (\Omega)Y_{\lambda' \mu'}^* (\Omega')e^{-i \vec{k}\cdot (\vec{\rho}-\vec{\rho}')} \, .
\eal
For a spherically symmetric impurity, i.e., in the limit of $U_{\lambda \neq 0 }(k) \to 0$, the potential reduces to
\be
\tilde{U}(\vec{\rho},\vec{\rho}') = \frac{1}{2\pi^2}\int d k \, k^2 \frac{|V_F(k)|^2}{\omega(k)} j_0 (k |\vec{\rho}-\vec{\rho}'|) \,,
\ee
which can also be obtained by tracing out the internal angular space, 
\be
\tilde{U}(\vec{\rho},\vec{\rho}') = \int d \Omega\, d \Omega' \tilde{U}(\vec{\rho},\vec{\rho}',\Omega,\Omega') \,.
\ee
This further simplifies to $\tilde{U}(\vec{\rho},\vec{\rho}') = \alpha_F/ (\sqrt{2} |\vec{\rho}-\vec{\rho}'|) $ for the Fr\"{o}hlich parameters which are given by a constant dispersion relation, $\omega(k) = \omega_0$, and the coupling strength,
\be
\label{frohlich_coup}
|V_F(k)| = \sqrt{2\sqrt{2}\pi \alpha_F/k^2} \, ,
\ee 
with $\alpha_F$ being the Fr\"{o}hlich particle-phonon coupling constant in units of $M=\omega_0=1$~\cite{Wu_86,Peeters_87,devreese2012physics}. 

\begin{figure}
  \centering
  \includegraphics[width=\linewidth]{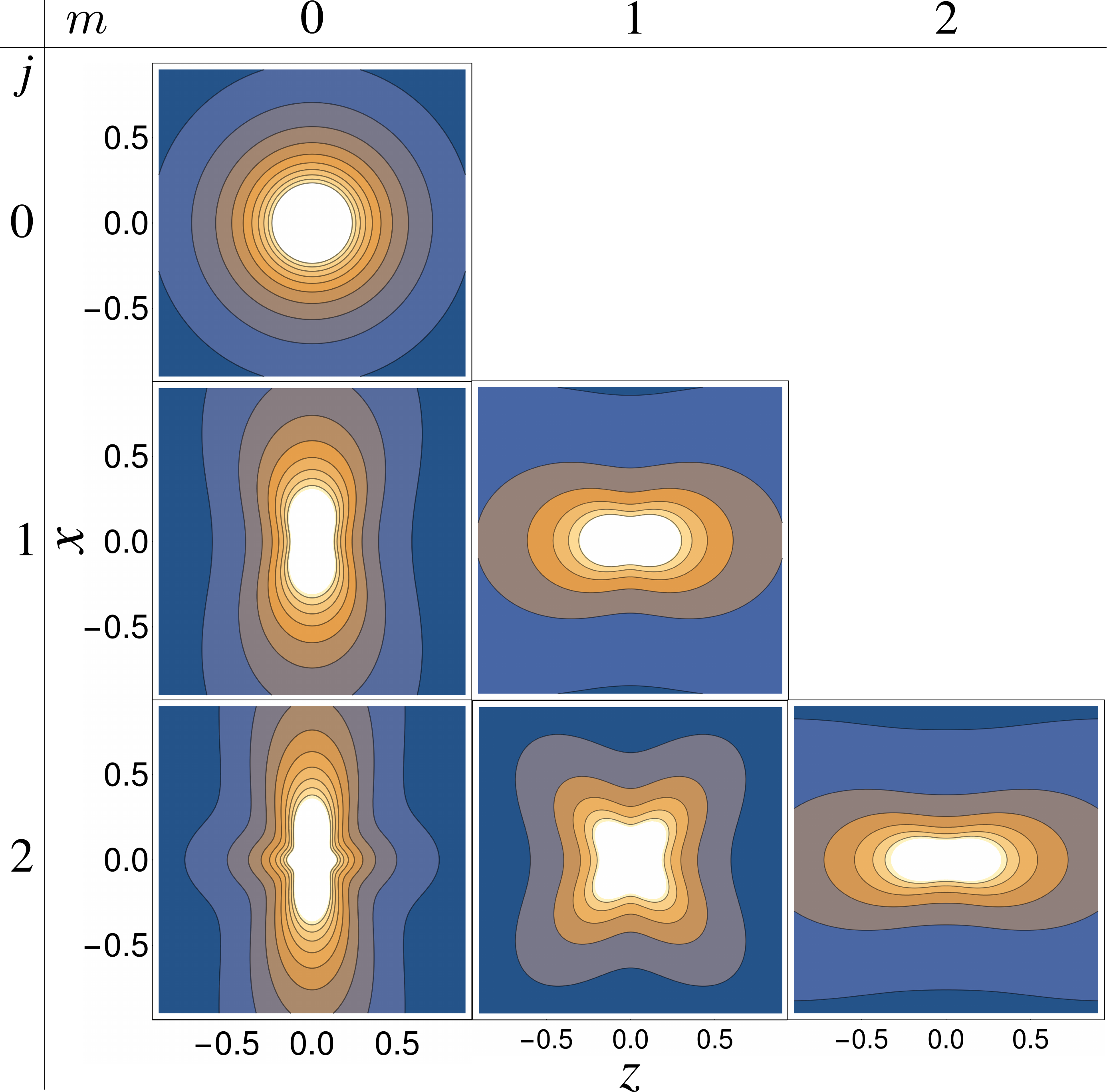}
 \caption{Self-induced potential of the impurity in a particular internal angular momenum state, $\ket{jm}$, where the Fr\"{o}hlich parameters~\eqref{frohlich_coup} are used.  The first figure with $j=0,m=0$ corresponds to the Fr\"{o}hlich polaron. See the text.}
 \label{pot}
\end{figure}

Moreover, if the impurity is in a definite internal angular momentum state, $\ket{jm}$, i.e., $\varphi_I(\vec{\rho},\Omega) \propto Y_{jm}(\Omega)$, from the Pekar functional~(\ref{pekar_functional}) we obtain:
\bal
\label{self_ind_pot}
& \nonumber \tilde{U}_{jm}(\vec{\rho},\vec{\rho}')  = \sum_{\lambda \lambda' l} i^{\lambda'-\lambda-l} \frac{(2\lambda+1)\sqrt{2l+1}}{\sqrt{4\pi}}\left[C_{\lambda 0, l 0}^{\lambda' 0}\right]^2 Y_{l,0}(\Omega_{\rho-\rho'}) \\
& \times C_{j 0, \lambda 0}^{j 0} C_{jm, \lambda 0}^{jm}C_{j 0, \lambda' 0}^{j 0}C_{jm, \lambda' 0}^{jm}\int d k \, \frac{U_\lambda(k)U_{\lambda'}^{*}(k)}{\omega(k)}j_l (k |\vec{\rho}-\vec{\rho}'| ) \, ,
\eal
which corresponds to an extended/anisotropic Fr\"{o}hlich impurity. As a model potential, we choose
\be
\label{cs}
U_\lambda (k) = V_F(k) \sqrt{\frac{2k^2}{\pi}} \frac{1}{\sqrt{2\lambda+1}} \, ,
\ee
where $U_0 (k)$ identifies the polaron coupling. Hereafter, Eq.~\eqref{cs}, with the previously defined Fr\"{o}hlich parameters~\eqref{frohlich_coup} and $B=1$,  will be considered as the default coupling.

For the coupling~(\ref{cs}), the self-induced potential~(\ref{self_ind_pot}) becomes proportional to $\int dk \, j_l (k |\vec{\rho}-\vec{\rho}'| ) Y_{l,0}(\Omega_{\rho-\rho'}) $, where the index $l$ is restricted by the Clebsch-Gordan coefficients $C_{\lambda 0, l 0}^{\lambda' 0} C_{j 0, \lambda 0}^{j 0} C_{j 0, \lambda' 0}^{j 0}$. In Fig.~\ref{pot}, we show the resulting potential for different $\ket{jm}$ states. The top panel corresponds to the Fr\"{o}hlich polaron, while the rest of the panels show distortions of the $1/r$ potential with increasing $j$ and $m$.

On the other hand, in the angulon limit, $M\to \infty$, the potential reads
\be
\label{angle_pot}
\tilde{U}(\Omega,\Omega') = \sum_{k \lambda} \frac{2\lambda+1}{4\pi} \frac{|U_\lambda(k)|^2}{\omega(k)} P_\lambda(\Omega-\Omega') \, ,
\ee
with $P_\lambda$ being the Legendre polynomial of degree $\lambda$. In fact, Eq.~(\ref{angle_pot}) can also be obtained for an impurity localized in the position space, $\varphi_I(\vec{\rho},\Omega) \propto \delta(\vec{\rho} -\vec{\rho}_0)$, which implies that the angulon is a translationally localized rotating polaron.

\subsection{Self-localization}

In general, the impurity wavefunction can be expanded in terms of spherical-wave states as
\be
\varphi_I (\vec{\rho},\Omega) = \int_0^\infty d p \sum_{l m_l j m_j} D_{l m_l, j m_j}(p) \, j_l (p \rho) \, Y_{l m_l} (\Omega_\rho) Y_{j m_j} (\Omega) \, .
\ee
Minimization of the energy with respect to the coefficients $D_{l m_l, j m_j}(p)$ identifies the corresponding Pekar energy. Here, as we consider a potential having the rotational symmetry with respect to the $z$-axis, see Eq.~(\ref{int_pot_bf}), we restrict the wavefunction to the $m_l = m_j = 0$ manifold. Therefore, we set $D_{l 0, j 0}(p) \equiv D_{l, j }(p)$ hereafter.

First of all, in the absence of the internal rotational degrees of freedom, i.e., for $j = 0$, the Fr\"{o}hlich polaron has a rotational symmetry in the translational space, and hence the impurity wavefunction can be modeled by the following radial Gaussian function~\cite{Devreese15}
\be
\varphi_I (\vec{\rho},\Omega) \to \frac{1}{\sqrt{4 \pi}} \left( \frac{\beta}{\pi} \right)^{3/4}e^{-\beta \rho^2/2} \, ,
\ee
with the variational parameter $\beta$. This indicates that the corresponding coefficient is given by
\be
D_{0,0} (p)=  \frac{2 p^2}{\pi} \int_0^\infty d \rho \, \rho^2 j_0 (p \rho) R_\beta (\rho) \,,
\ee
where 
\be
R_\beta (\rho) = \sqrt{\frac{4\beta^{3/2}}{\sqrt{\pi}}}e^{-\beta \rho^2/2} \, ,
\ee
and we used the relation $\int_0^\infty d\rho \, \rho^2 j_l (\rho p) j_l (\rho p') = \pi \delta(p-p')/(2 p^2)$. 

\begin{figure}
  \centering
  \includegraphics[width=0.9\linewidth]{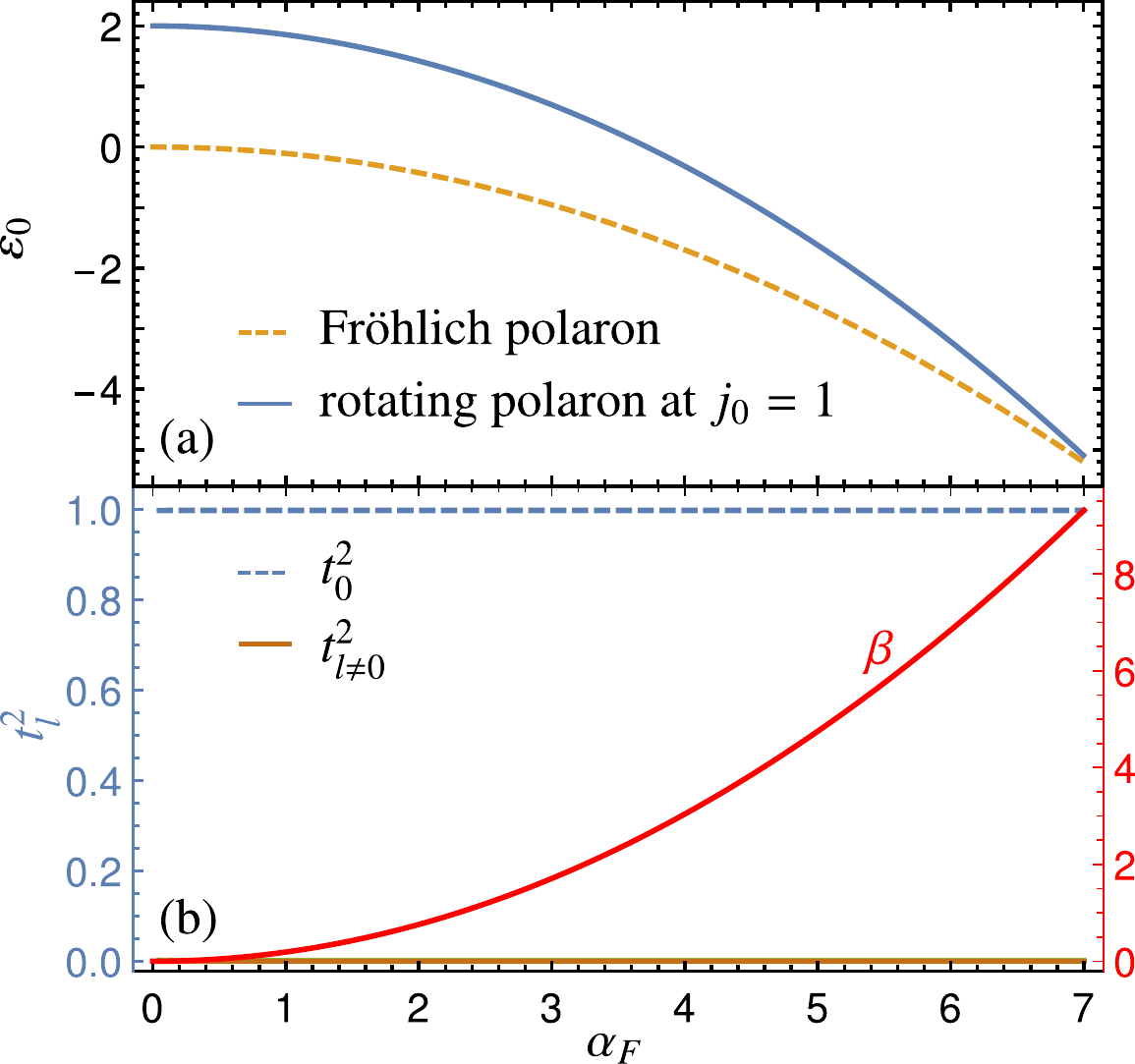}
 \caption{(a) Comparison of the Pekar energy of the Fr\"{o}hlich polaron (dashed yellow curve), and that of the rotating polaron in the angular momentum state $j_0=1$, as  given by Eq.~\eqref{pekar_ansatz_gaussian} or Eq.~\eqref{pekar_ansatz_j} (blue curve), as a function of the coupling constant, $\alpha_F$. (b) The variational parameters, $t_l$ and $\beta$, of the ansatz~\eqref{pekar_ansatz_j} as a function of the coupling constant. The angular momentum cutoff is $l_\text{max} =10$. See the text.}
 \label{pr}
\end{figure}

Next, in order to analyze how the internal rotational degrees of freedom affect the translational part of the impurity wavefunction, we consider a simple case where the impurity has a definite rotational angular momentum state, $j_0$. Accordingly, we write down the following ansatz, with the variational parameters, $\beta$ and $\gamma$,
\be
\label{pekar_ansatz_gaussian}
\varphi_I (\vec{\rho},\Omega) = Y_{j_0 0} (\Omega) \sqrt{\frac{\beta \sqrt{\beta+\gamma}}{\pi^{3/2}}}e^{-\beta \rho^2/2} e^{-\gamma \rho^2 \cos^2(\theta_k) /2} \, .
\ee
The corresponding self-induced potentials due to the internal rotational state can be seen in Fig.~\ref{pot}. Then, the Pekar energy for the coupling defined in Eq.~\eqref{cs} can be written as
\bal
\label{pekar_en_gaussian}
\nonumber \varepsilon_0 \left[\beta, \gamma \right] & = j_0 (j_0 + 1) + \frac{3\beta +\gamma}{4} - \alpha_F \sqrt{\frac{2}{\pi}}  \sum_{\lambda \lambda'}  \left[ C_{j_0 0, \lambda 0}^{j_0 0}\right]^2 \left[ C_{j_0 0, \lambda' 0}^{j_0 0}\right]^2 \\
& \times i^{\lambda'-\lambda} \int d \Omega_k \, Y_{\lambda 0} (\Omega_k) Y_{\lambda' 0} (\Omega_k) \sqrt{\frac{\beta( \beta+  \gamma) }{2 \beta + \gamma-\gamma  \cos (2 \theta_k ) }} \, .
\eal
Apart from very small values of the coupling $\alpha_F$ (see the discussion below), numerical minimization of Eq.~\eqref{pekar_en_gaussian} with respect to the parameter $\gamma$ yields 
\be
\argmin_{\gamma} \varepsilon_0 \left[\beta, \gamma \right] = 0 \, .
\ee
This further simplifies the Pekar energy to
\be
\label{pekar_energy_gaussian}
\varepsilon_0 = j_0 (j_0 + 1)- \frac{\alpha_F^2}{3\pi}\left( \sum_\lambda \left[ C_{j_0 0, \lambda 0}^{j_0 0}\right]^4\right)^2 \, .
\ee
In the absence of  internal rotational degrees of freedom, i.e., for $j_0 \to 0$, the Pekar energy corresponds to the standard Fr\"{o}hlich polaron, $\varepsilon_0 = - \alpha_F^2 /(3 \pi)$, which is shown in Fig.~\ref{pr}~(a) with the dashed yellow curve. In the same figure, we show the respective Pekar energy in the case of $j_0 = 1$ by the blue curve. Since the wavefunction is in a definite rotational angular momentum state, the energy at $\alpha_F =0$ starts at $2B$. Its qualitative behavior, however, is similar to the Fr\"{o}hlich polaron.

As the second ansatz, we consider
\be
\label{pekar_ansatz_j}
 \varphi_I (\vec{\rho},\Omega) =  Y_{j_0,0}(\Omega) R_\beta (\rho) \sum_l^{l_\text{max}} t_{l} Y_{l,0}(\Omega_\rho)  \, ,
\ee
with $l_\text{max}$ being an angular momentum cutoff. If we plot the translational angular variational parameters, $t_l$, which minimize the respective Pekar energy, we observe that they are given by $t_l \approx \delta_{l0}$ for all values of the coupling constant, see Fig.~\ref{pr}~(b).

This shows that both approaches, Eqs.~\eqref{pekar_ansatz_j}  and \eqref{pekar_ansatz_gaussian}, lead to the same wavefunction and the same Pekar energy, Eq.~\eqref{pekar_energy_gaussian}. In particular, both variational wavefunctions reveal that although the self-induced potential is distorted from the $1/r$ shape due to the nonzero internal angular momentum state, the translational part of the wavefunction remains radially symmetric as in the Fr\"{o}hlich case. Heuristically this is explained as follows: in the  Fr\"{o}hlich case one finds that the optimal value of the variational parameter is given by $\beta = 4 \alpha_F^2/(9 \pi)$. Then, almost the entire Gaussian wavefunction is located within a sphere centred at the origin with a radius $\sqrt{9 \pi/(8 \alpha_F^2)}$ of three times the standard deviation. For large values of $\alpha_F$, the electron is strongly localized and only sees the potential in the near vicinity of the origin.
While the self-induced potential from Fig.~\ref{pot} gets distorted by the internal angular state on a scale of order one, it is still spherically symmetric near the origin. Thus if $\alpha_F$ is large and the electron is  sharply localized, the shape of its wavefunction does not change since  it effectively experiences a radial potential.  In contrary, for small values of $\alpha_F$,  which is outside of the scope of the Pekar ansatz, the electron wavefunction is more delocalized and is modified by a nonzero value of the parameter $\gamma$.

Based on the  discussion above, for the most general case, we make an ansatz, $D_{l, j}(p) = D_{l,0} (p) \, d_{l, j}$, such that the radial part of the wavefunction, $ D_{l,0} (p)$, is decoupled from its angular part, $d_{l, j}$. This suggests us to write a trial wavefunction for the rotating polaron in the form of
\be
\label{pekar_ansatz}
\varphi_I (\vec{\rho},\Omega) = R_\beta (\rho) \sum_{l j}^{l_\text{max}, j_\text{max}} d_{l, j} Y_{l 0} (\Omega_\rho)Y_{j 0} (\Omega) \, ,
\ee
with $l_\text{max}, j_\text{max}$ being the angular momentum cutoff. The remaining parameters, $\beta$ and $d_{l,j}$, with the normalization condition $\sum_{lj} |d_{l,j}|^2 =1$, are considered as variational parameters. 

Using Eq.~\eqref{pekar_ansatz}, the Pekar energy for the coupling defined in Eq.~(\ref{cs}) can be written as
\be
\varepsilon_0 \left[\beta, \{d_{l,j} \} \right] = \beta \, \mathcal{C}_1 \left[\{d_{l,j} \} \right]  - \sqrt{\beta} \, \alpha_F \, \mathcal{C}_2 \left[\{d_{l,j} \} \right] + \mathcal{C}_3 \left[\{d_{l,j} \} \right] \, ,
\ee
where
\bal
\nonumber \mathcal{C}_1 \left[\{d_{l,j} \} \right] & = \left(\frac{3}{4} + \sum_{lj} |d_{l,j}|^2  \, l(l+1) \right) \, , \\
\mathcal{C}_2 \left[\{d_{l,j} \} \right] & = \left. \sum_{\vec{k}} \frac{|\tilde{V}(\vec{k})|^2}{\omega(k)} \right|_{\beta=1, \alpha_F = 1} \, , \\
\nonumber \mathcal{C}_3 \left[\{d_{l,j} \} \right] & =\sum_{lj} |d_{l,j}|^2  \, j(j+1) \, .
\eal
After minimization of the energy with respect to the parameter $\beta$, we find
\be
\beta = \left( \frac{\alpha_F \, \mathcal{C}_2 \left[\{d_{l,j} \} \right]}{2\, \mathcal{C}_1 \left[\{d_{l,j} \} \right]} \right)^2 \, .
\ee
As a result of this, the Pekar energy reads
\be
\label{pekar_general_d}
\varepsilon_0 \left[ \{d_{l,j} \} \right] = - \frac{\alpha_F^2 \, \mathcal{C}_2 \left[\{d_{l,j} \} \right]^2}{4 \, \mathcal{C}_1 \left[\{d_{l,j} \} \right]} + \mathcal{C}_3 \left[\{d_{l,j} \} \right] \, .
\ee
Naturally, the limiting case of $d_{l,j} =\delta_{l 0} \delta_{j 0}$ corresponds to the standard Fr\"{o}hlich polaron.

\begin{figure}
  \centering
  \includegraphics[width=0.9\linewidth]{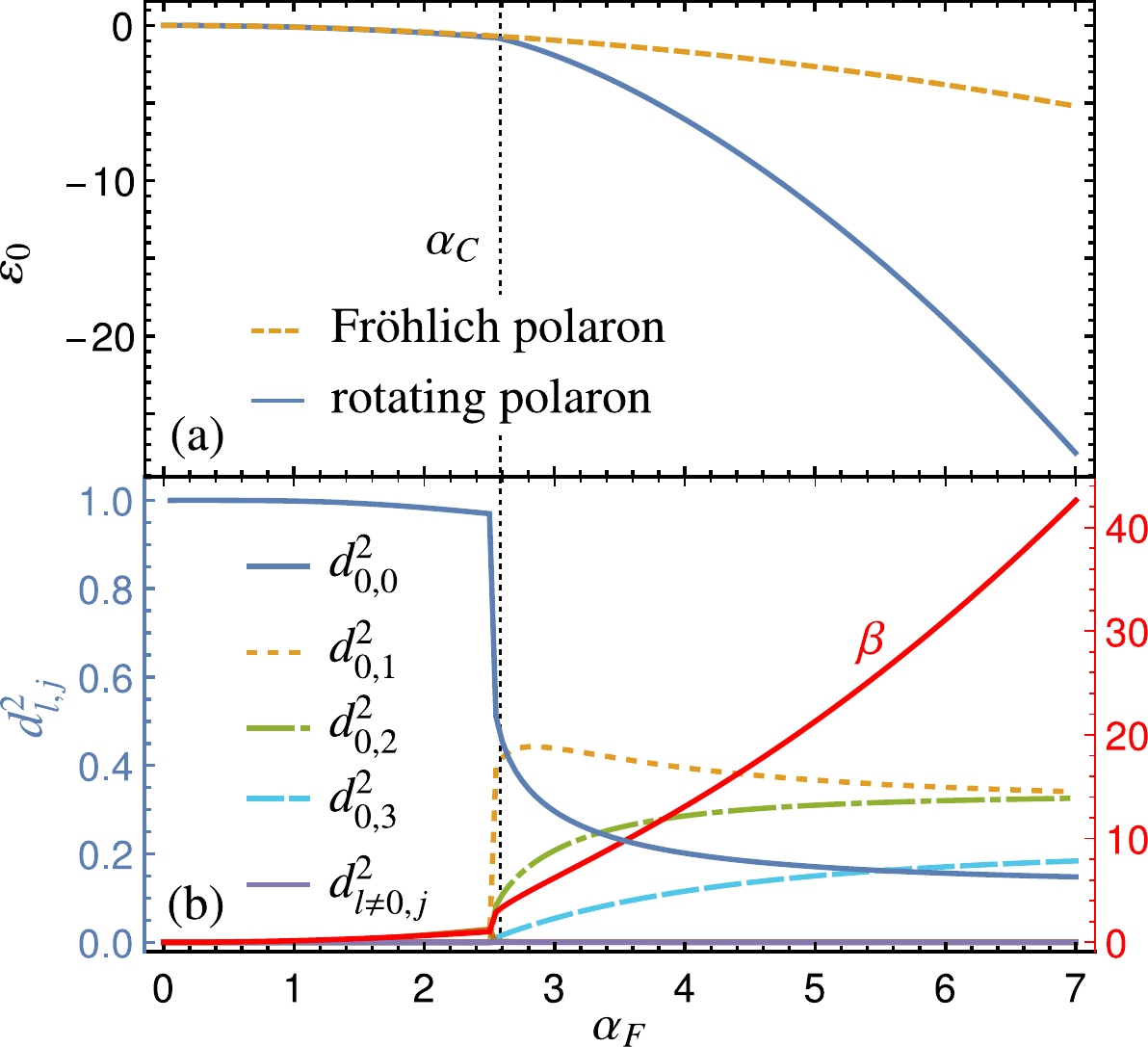}
 \caption{(a) Comparison of the Pekar energy of the Fr\"{o}hlich polaron (dashed yellow curve), and that of the rotating polaron given by Eq.~(\ref{pekar_ansatz}) with the angular momentum cutoff $l_\text{max}, j_\text{max} = 3$ (blue curve) as a function of the coupling constant, $\alpha_F$. After a critical coupling, $\alpha_C \approx 2.5$ (dotted vertical line), the energy features a kink. (b) Variational parameters $d_{l,j}$ and $\beta$ as a function of the coupling constant. See the text.}
 \label{ptr}
\end{figure}

Now, let us consider the general state~(\ref{pekar_ansatz}) with the angular momentum cutoff $l_\text{max}, j_\text{max} = 3$. The resulting energy is shown in Fig.~\ref{ptr}~(a). A very first observation is that after a certain coupling constant, $\alpha_C \approx 2.5$, the energy sharply decreases. In fact, the derivative of the energy with respect to the coupling constant $\alpha_F$ features a discontinuity at the critical coupling strength resulting in a `kink' in  the blue line in Fig.~\ref{ptr}~(a). Such a behavior of the energy corresponds to the phenomenon of a self-localization transition in the vicinity of the kink~\cite{landau1933uber,Li_2017}.

The possibility of a self-localization transition was discussed for the first time in the seminal papers of Landau and Pekar~\cite{landau1933uber,pekar1946local},  and attracted a lot of theorists' attention in various polaron models after Fr\"{o}hlich introduced a microscopic model describing polaron~\cite{frohlich1954electrons}. However, whether the transition really exists or arises solely due to the applied approximations  is a long-standing and highly debated problem. For example, although several theories predicted the existence of a self-localization transition for the Fr\"{o}hlich polaron~\cite{gross1959analytical,PMatz_71,manka1978first,lepine1979mean, Luttinger_80, sumi1977phase,peeters1982phase,tokuda1981optical}, later it has been shown to be the artefact of the theoretical approach~\cite{Fisher_86,Gerlach_91,Mishchenko_00,peeters1982existence, feranchuk2005new}.

For the Fr\"{o}hlich polaron, at the level of the Pekar approximation, the electron wavefunction is localized at any coupling strength $\alpha_F$. In fact, it has been proven~\cite{lieb1977existence} that there exist a unique minimizer (up to translations) for the Pekar energy functional  such that there is no self-localization transition at finite $\alpha_F$. However,  it was recently shown that an angular self-localization transition takes place in the angulon problem already at the Pekar level~\cite{Li_2017} .

In order to analyze the observed kink in the energy, in Fig.~\ref{ptr}~(b) we plot the variational parameters, $d_{l,j}$ and $\beta$, as a function of the coupling constant, $\alpha_F$. First of all, in the vicinity of the critical coupling, the value of $\beta$ jumps, and after the critical coupling, it swiftly increases. Furthermore, for small values of the coupling constant, $\alpha_F < \alpha_C$, the angular variational parameters are given by $d_{l,j} \approx \delta_{l 0} \delta_{j 0}$. This behavior  corresponds to the standard Fr\"{o}hlich polaron. On the other hand, after the critical value, these parameters become $d_{l,j} \propto \delta_{l0}$. In other words, while the internal angular states get superposed, the translational angular states with $l\neq 0$ vanish for all values of $\alpha_F$ such that the rotational part of the impurity wavefunction decouples from the rest of it. This suggests that we can consider the following trial wavefunction,
\be
\label{pekar_ansatz_l}
 \varphi_I (\vec{\rho},\Omega) = R_\beta (\rho) Y_{0,0}(\Omega_\rho)\sum_j^{j_\text{max}} r_{j} Y_{r,0}(\Omega) \, ,
\ee
which allows to substantially simplify the variational calculation and to derive a more transparent model.

It follows from Eq.~(\ref{pekar_general_d}) that the corresponding Pekar energy in this case is given by
\be
\label{pekar_anal}
\varepsilon_0 \left[\{r_j \}_0^{j_\text{max}} \right]=  - \frac{\alpha_F^2}{3\pi} \mathcal{B}\left[\{r_j \}_0^{j_\text{max}} \right]^2 + \sum_j |r_j|^2 j(j+1)\, ,
\ee
where 
\be
\mathcal{B}\left[\{r_j \}_0^{j_\text{max}} \right] = \sum_\lambda \left| \sum_{j j'} r_j r_{j'}^* \sqrt{\frac{2j'+1}{2j+1}} \left[ C_{j'0,\lambda 0}^{j0}\right]^2 \right|^2 \, .
\ee

\begin{figure}
  \centering
  \includegraphics[width=0.9\linewidth]{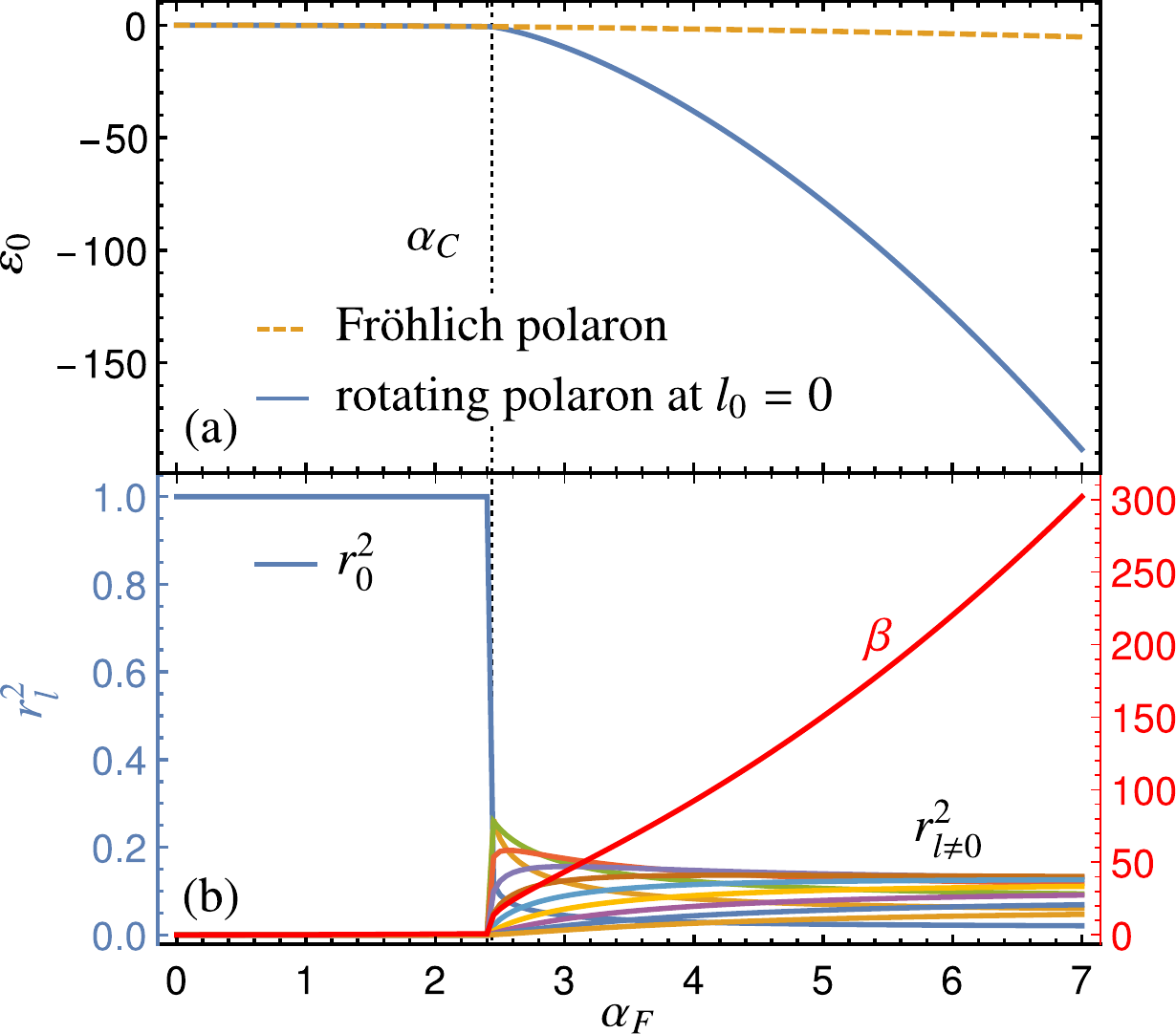}
 \caption{(a) Comparison of the Pekar energy of the Fr\"{o}hlich polaron (dashed yellow curve), and that of the rotating polaron given by Eq.~(\ref{pekar_ansatz_l}) with the angular momentum cutoff $j_\text{max} = 10$ (blue curve) as a function of the coupling constant, $\alpha_F$. (b) Variational parameters, $r_{j}$ and $\beta$, as a function of the coupling constant. After the critical coupling, $\alpha_C \approx 2.4$ (dotted vertical line), the impurity wavefunction becomes localized in both the rotational and translational space. See the text.}
 \label{pt}
\end{figure}

For very small values of the coupling constant, 
the argument of the minimum value of the Pekar energy yields
\be
\label{weak_a}
\argmin_{r_j} \varepsilon_0 \left[\{r_j \}_0^{j_\text{max}} \right] = \argmin_{r_j }\sum_j |r_j|^2 j(j+1) = \{\delta_{j0}\} \, .
\ee
Therefore, the minimum value of the energy is given by $\min_{r_j} \varepsilon_0 = -\alpha_F^2 / (3 \pi)$, which corresponds to the Fr\"{o}hlich case, as  discussed above. However, for very large values of $\alpha_F$, the argument of the minimum value of the energy reads:
\be
\label{strong_a}
\argmin_{r_j} \varepsilon_0 \left[\{r_j \}_0^{j_\text{max}} \right] = \argmax_{r_j } \mathcal{B}\left[\{r_j \}_0^{j_\text{max}} \right]^2 = \left\{\frac{1}{\sqrt{j_\text{max}+1}}\right\} \, .
\ee

In Fig.~\ref{pt}~(a), the Pekar energy~(\ref{pekar_anal}) is shown for the angular momentum cutoff $j_\text{max} =10$, see the blue curve. The behavior of the energy is very similar to the previous case of Fig.~\ref{ptr}~(a). After the critical coupling constant, $\alpha_C \approx 2.4$, the energy features a kink. Fig.~\ref{pt}~(b) shows the variational parameters, $r_j$ and $\beta$. As we discuss above, for $\alpha_F < \alpha_C$, the rotational internal state is given by $r_j = \delta_{j 0}$. For $\alpha_F > \alpha_C$, on the other hand, the rotational wavefunction is almost an equal-weight superposition of all angular states, which yields $r_j \approx 1/\sqrt{j_\text{max}+1}$. Therefore, we deduce that the internal rotational motion of the impurity, which is delocalized for $\alpha_F < \alpha_C$ becomes localized in the space of angles for  $\alpha_F > \alpha_C$. This result has been previously found in Ref.~\cite{Li_2017}. It is this localization transition of the internal degree of freedom and the corresponding change in the self-induced potential that causes strongly enhanced self-localization of the translational degree of freedom.

\section{Conclusions} \label{sec_conc}

To summarize, we derived a novel Hamiltonian describing linear molecules (or other impurities with a translational and an internal rotational degree of freedom) immersed in a bosonic bath, e.g.\ liquid Helium or a BEC of cold atoms. While such a Hamiltonian represents a hybrid between the well-established polaron and the recently introduced angulon quasiparticles, we expect it to go beyond just the sum of its parts and to host novel impurity physics.

Here we undertook the first steps in the study of this Hamiltonian by considering the solutions in the weak- and strong-coupling regimes. In weak coupling, we analyzed the spectrum based on a single-phonon variational state describing a rotating impurity with a fixed linear momentum. We found that long-lived quasiparticle states (`rotating polarons') strongly  depend  on the total linear momentum of the combined impurity-bath system. In particular, the instability regions where the quasiparticle picture breaks down become substantially larger with growing linear momentum. The strong-coupling regime has been analyzed using the Pekar-type mean-field ansatz. First, we have demonstrated that, in our parameter regime, the translational part of the rotating polaron is spherically symmetric. This is the case even for higher internal angular momentum states which give rise to a non-spherically symmetric self-induced potential. Then it has been shown that the self-localization transition previously discussed for the angulon~\cite{Li_2017}, also takes place in the internal rotational degree of freedom of the rotating polaron. Moreover, such rotational localization triggers a transition of enhanced localization in the translational degree of freedom. This reveals that the rotational localization transition, which is probably challenging to detect directly in experiment, can be accessed by measuring the spatial extension of the rotating polaron. Our findings shed the first light on the interesting and rich physics of the rotating polaron, and can be applicable to a variety of systems, from Rydberg atoms~\cite{Camargo_18} to electronic impurities with angular momentum in solids~\cite{Levinsen_12,das1993study}.

\begin{acknowledgments}

We are grateful to G. Bighin and X. Li for valuable discussions. E.Y. and B.M. acknowledge financial support received from the People Programme (Marie Curie Actions) of the European Union's Seventh Framework Programme (FP7/2007-2013) under REA grant agreement No. [291734]. A.D. and N.L. acknowledge support from the European Research Council (ERC) under the European Union’s Horizon 2020 research and innovation program (Grant Agreement No. 694227). M.L. acknowledges support from the Austrian Science Fund (FWF), under project No. P29902-N27.
\end{acknowledgments}

\appendix
\section{Detailed derivation of the Hamiltonian} \label{appendix}

Here we derive the Hamiltonian, $\hat{H}_{\rm int}$, describing the interactions between an impurity and a many-body environment in the laboratory reference frame. 

First, we consider a mobile molecular impurity in a weakly interacting atomic BEC, where the coordinate of an atom interacting with the impurity is given by ${\bf R} =(R,\Theta_R,\Phi_R)$, ${\bf r} =(r,\theta_r,\phi_r)$, and ${\bf r}' =(r',\theta'_{r'},\phi'_{r'})$ in the laboratory, molecular body-fixed, and molecular center-of-mass (CM) frames of reference, respectively, cf. Fig.~\ref{diagram}.  The interaction part of the Hamiltonian can be written as follows \cite{LemSchmidtChapter}:
\begin{equation}\label{Eq-1}
\hat{H}_{\rm int} = \sum\limits_{\bf k,q} e^{-i{\bf q}\cdot\hat{\boldsymbol\rho}} V({\bf q},\hat{\theta},\hat{\phi}) ~\hat{a}^\dag_{\bf k+q} \hat{a}_{\bf k} \, .
\end{equation}
Here the operator $\hat{\boldsymbol \rho}\equiv (\hat{\rho},{\hat\theta}_\rho,\hat{\phi}_\rho)$ measures an instantaneous position of the molecule's CM with respect to the laboratory frame, $(\hat{\theta}, \hat{\phi})$ are the angle operators determining the molecule's instantaneous orientation in space, $\hat{a}^\dag_{\bf k}$ and $\hat{a}_{\bf k}$ are the bosonic creation and annihilation operators, respectively. $V({\bf q},\hat{\theta},\hat{\phi}) = \mathcal{F}[D(\hat{\phi}, \hat{\theta}, 0) V({\bf r})]$ is the potential in Fourier space after rotating the body-fixed coordinates into the CM frame by using the Wigner rotation operator $D(\hat{\phi}, \hat{\theta}, \hat{\gamma})$. In order to derive the expression for $V({\bf q},\hat{\theta},\hat{\phi})$, we first expand the interaction potential between a molecule and an atom in the molecular body-fixed frame, $V(\vec{r})$, over the spherical harmonics as
\be
\label{int_pot_bf}
V({\bf r}) = \sum_\alpha V_\alpha(r) Y_{\alpha 0}(\theta_r,\phi_r) \, .
\ee
By using the Wigner rotation matrices, $Y_{\alpha 0}(\theta_r,\phi_r) = \sum_\gamma D_{\gamma 0}^\lambda (\hat{\phi},\hat{\theta},0)Y_{\alpha \gamma}(\theta'_{r'},\phi'_{r'})$, we write the interaction potential~(\ref{int_pot_bf}) in the CM-frame as
\be 
V(\vec{r}',\hat{\theta},\hat{\phi}) = \sum_{\alpha \gamma} V_\alpha(r) Y_{\alpha \gamma}(\theta'_{r'},\phi'_{r'}) Y_{\alpha\gamma}(\hat{\theta},\hat{\phi}) \, .
\ee
The potential in Fourier space is then given by
\bal
\label{Eq-2}
V({\bf q},\hat{\theta},\hat{\phi}) & = \int d^3 r' \, V(\vec{r}',\hat{\theta},\hat{\phi}) e^{-i \vec{q} \cdot \vec{r}'} \\
\nonumber & = \sum\limits_{\alpha\gamma} (2\pi)^{3/2} i^{-\alpha} \tilde{V}_\alpha(q) Y_{\alpha\gamma}(\theta_q,\phi_q) Y_{\alpha\gamma}(\hat{\theta},\hat{\phi}) \, ,
\eal 
with $\tilde{V}_\alpha (q) = 2^{3/2}(2\alpha+1)^{-1/2} \int_0^\infty dr~ r^2 V_\alpha(r) j_\alpha(q r)$ being the spherical Fourier transform of the expansion coefficients $V_{\alpha}(r)$. In Eq.~(\ref{Eq-2}) we have used the plane wave expansion, $e^{-i{\bf q}\cdot\hat{\boldsymbol\rho}} = 4\pi \sum_{\ell\delta} i^{-\ell} j_\ell(q \hat{\rho}) Y^*_{\ell\delta}(\hat \theta_\rho,\hat \phi_\rho) Y_{\ell\delta}(\theta_q,\phi_q)$, and $r' = r$.

In the case of a weakly interacting BEC, applying the Bogoliubov approximation to the Hamiltonian \eqref{Eq-1} gives:  
\begin{equation}
\hat{H}_{\rm int} = \sqrt{n} \sum\limits_{\bf k} \sqrt{\frac{\epsilon(k)}{\omega(k)}} e^{-i{\bf k}\cdot\hat{\boldsymbol\rho}} V({\bf k},\hat{\theta},\hat{\phi}) \hat{b}^\dag_{\bf k} + \text{H.c.}\,
\end{equation}
Now using Eq.~\eqref{Eq-2} and the spherical expansion of the boson operators, $\hat{b}^\dag_{\bf k} = (2\pi)^{3/2} k^{-1} ~ \sum_{\lambda\mu} i^\lambda Y^*_{\lambda \mu}(\theta_k,\phi_k)~ \hat{b}^\dag_{k\lambda \mu}$, we obtain
\begin{equation}\label{Eq-4}
\hat{H}_{\rm int} =  \sum\limits_{ k \lambda \mu} \sum\limits_{\ell \delta \alpha\gamma} \mathcal{U}_{\ell\alpha\lambda}^{\delta\gamma\mu}(k)~j_\ell(k \hat{\rho})  ~Y_{\ell \delta}^*(\hat{\theta}_\rho,\hat{\phi}_\rho)~Y_{\alpha \gamma}^*(\hat{\theta},\hat{\phi}) ~\hat{b}^\dag_{k\lambda \mu} +  \text{H.c.} \, ,
\end{equation}
where
\bal
\mathcal{U}_{\ell\alpha\lambda}^{\delta\gamma\mu}(k) & = U_\alpha (k) \sqrt{4\pi(2\alpha+1)(2l+1)/(2\lambda+1)} \\
\nonumber & \times i^{\lambda-\alpha-l} C^{\lambda 0}_{\alpha 0, l 0} C^{\lambda \mu}_{\alpha \gamma, l \delta} \, ,
\eal
with
\be
U_\lambda (k)=  \sqrt{\frac{8 n k^2 \epsilon (k)}{\omega (k) (2\lambda+1)}} \int_0^\infty dr r^2 V_\lambda (r) j_\lambda (k r) \, ,
\ee
and we have used that~\cite{Varshalovich}
\bal
\nonumber & \int d \Omega_k ~ Y_{\ell\delta}(\Omega_k) Y_{\alpha\gamma}(\Omega_k) Y^*_{\lambda\mu}(\Omega_k)\\
& = \sqrt{\frac{(2\ell+1)(2\alpha+1)}{4\pi(2\lambda+1)}} C^{\lambda 0}_{\ell 0,\alpha 0} C^{\lambda \mu}_{\ell \delta,\alpha \gamma} \, . 
\eal


\bibliography{ap.bib}

\end{document}